\documentclass[runningheads]{llncs}
\usepackage{makecell}
\usepackage{listings}
\usepackage{graphicx}
\usepackage{xcolor}
\usepackage{comment}
\usepackage{algorithm}
\usepackage{hyperref}
\usepackage[noend]{algpseudocode}
\usepackage[inline]{enumitem}
\usepackage{xcolor}
\usepackage{listings}
\usepackage{tcolorbox}
\usepackage{amsmath}
\usepackage{amssymb}
\usepackage{url}
\usepackage{xspace,framed}
\definecolor{codegreen}{rgb}{0,0.6,0}
\definecolor{codegray}{rgb}{0.5,0.5,0.5}
\definecolor{codepurple}{rgb}{0.58,0,0.82}
\definecolor{backcolour}{rgb}{0.95,0.95,0.92}

\lstdefinestyle{mystyle}{
    backgroundcolor=\color{backcolour},   
    commentstyle=\color{codegreen},
    keywordstyle=\color{magenta},
    numberstyle=\tiny\color{codegray},
    stringstyle=\color{codepurple},
    basicstyle=\ttfamily\footnotesize,
    breakatwhitespace=false,         
    breaklines=true,                 
    captionpos=b,                    
    keepspaces=true,                 
    numbers=left,                    
    numbersep=5pt,                  
    showspaces=false,                
    showstringspaces=false,
    showtabs=false,                  
    tabsize=2
}

\newcommand\tool{{\sf EBF}\xspace}

\lstdefinestyle{mystyle}{
  backgroundcolor=\color{backcolour},   commentstyle=\color{codegreen},
  keywordstyle=\color{magenta},
  numberstyle=\tiny\color{codegray},
  stringstyle=\color{codepurple},
  basicstyle=\ttfamily\footnotesize,
  breakatwhitespace=false,         
  breaklines=true,                 
  captionpos=b,                    
  keepspaces=true,                 
  numbers=left,                    
  numbersep=5pt,                  
  showspaces=false,                
  showstringspaces=false,
  showtabs=false,                  
  tabsize=2
}

\begin{document}
\title{Finding Security Vulnerabilities in IoT Cryptographic Protocol and Concurrent Implementations \thanks{Supported by EPSRC grants  EP/T026995/1 and EP/V000497/1. The first author acknowledges the scholarship she is receiving from King Faisal University (KFU).}}
\titlerunning{EBF: A Hybrid Verification Tool for IoT Cryptographic Protocols }
%\textcolor{red}{MM: This note breaks the double-blind paper reviewing process. I suggest we remove it when we submit the paper for review. We should only include it in the camera-ready version or the pre-print version.}
\author{ }
\institute{ }

\author{Fatimah Aljaafari\inst{1} \and
Rafael Menezes\inst{2} \and
Mustafa Mustafa\inst{1}\and
Lucas Cordeiro\inst{1} }

\authorrunning{Aljaafari et al.}

\institute{The University of Manchester, UK \and
Federal University of Amazonas, Brazil}

\maketitle              % typeset the header of the contribution

\begin{abstract}
Internet of Things (IoT) consists of a large number of devices connected through a network, which exchange a high volume of data, thereby posing new \textit{security}, \textit{privacy}, and \textit{trust} issues. One way to address these issues is ensuring \textit{data confidentiality} using lightweight encryption algorithms for IoT protocols. However, the design and implementation of such protocols is an error-prone task; flaws in the implementation can lead to devastating security vulnerabilities. Here we propose a new verification approach named Encryption-BMC and Fuzzing (\tool), which combines Bounded Model Checking (BMC) and Fuzzing techniques to check for security vulnerabilities that arise from concurrent implementations of cyrptographic protocols, which include data race, thread leak, arithmetic overflow, and memory safety. \tool models IoT protocols as a client and server using POSIX threads, thereby simulating both entities' communication. It also employs static and dynamic verification to cover the system's state-space exhaustively. We evaluate \tool against three benchmarks. First, we use the concurrency benchmark from SV-COMP and show that it outperforms other state-of-the-art tools such as \textit{ESBMC}, \textit{AFL}, \textit{Lazy-CSeq}, and \textit{TSAN} with respect to bug finding. Second, we evaluate an open-source implementation called WolfMQTT. It is an MQTT client implementation that uses the WolfSSL library. We show that \tool detects a data race bug, which other approaches are unable to find. Third, to show the effectiveness of \tool, we replicate some known vulnerabilities in OpenSSL and CyaSSL (lately WolfSSL) libraries. \tool can detect the bugs in minimum time.
%\keywords{Cryptographic protocols  \and Concurrency  \and Fuzzing \and Bounded Model Checking.}
\end{abstract}

%----------------------------------------
\section{Introduction}
%----------------------------------------
%\subsection{\textcolor{red}{Motivation}}
An Internet of Things (IoT) system usually comprises a large number of smart devices and objects, such as RFID tags, sensors, actuators, and smartphones, which communicate with each other (usually via Wifi, Bluetooth, and RFID) with minimum human interactions~\cite{2}. 
IoT covers different areas and applications, such as smart homes, cities, and health care~\cite{7}. According to Maayan ~\cite{80}, from 2020 to 2030, the number of IoT devices is expected to grow from $75$ billion to more than $100$ billion, the upgrade from 4G to 5G playing an important part in this growth. This large number of devices will create a massive and complex network with an exceedingly high volume of data communicated over it~\cite{balte2015security,12}. The existence of such a network of connected devices will inevitably pose new \textit{security}, \textit{privacy} and \textit{trust} issues that can put users at high risk~\cite{2}.

To address these issues, achieving \textit{data confidentiality} is paramount. A natural way to protect the data in transit is by designing bespoke lightweight encryption algorithms for IoT devices~\cite{14}. Due to limitations in IoT devices such as limited power supply, low memory, and low processing speed, lightweight encryption algorithms have been developed~\cite{29,30}. There are many established libraries for IoT cryptography, a good representative of them is WolfSSL, which is the focus of this paper. WolfSSL is a library targeted at resource-constrained devices due to its small size, speed, and feature set~\cite{94}. It provides lightweight implementations that support TLS/SSL, which supports various cryptographic algorithms, including lightweight encryption algorithms.
However, designing and implementing such algorithms for general cryptographic protocols is an error-prone task; flaws in the implementation can lead to devastating security vulnerabilities~\cite{nebbione2020security}.

%\subsection{\textcolor{red}{existing literature}}

Generally, there exist various techniques for finding security vulnerabilities~\cite{16,50}. One of them is Bounded Model Checking (BMC)~\cite{Biere09} which searches for violation in bounded executions of length $k$. If no bug is detected, then $k$ is increased until a bug is detected, the verification problem becomes intractable, or a pre-set upper bound is reached. Some examples of BMC tools include C Bounded Model Checker \textit{CBMC}~\cite{kroening2014cbmc} and Efficient SMT-based Bounded Model Checker \textit{ESBMC}~\cite{78}. Another popular technique is fuzzing~\cite{50}. It is an automated software testing technique that involves providing invalid values as inputs to a program. Then, the system behavior is checked for abnormalities, such as crashes or failures~\cite{27}. American Fuzzing Lop (\textit{AFL})~\cite{79} and \textit{LibFuzzer}~\cite{libfuzzer} are some of the state-of-the-art tools that implement fuzzing.

%\subsection{\textcolor{red}{challenges }}

 Recent years have seen a real development in software verification of cryptographic protocols and concurrent applications, as witnessed by the development of different tools using either BMC or fuzzing techniques~\cite{26}. However, there still exists a need for further development of these tools. BMC alone is inadequate for efficiently achieving high-path coverage, especially for multi-threaded programs. BMC struggles in reaching deep parts of the code because of the state-space explosion issue and its dependency on Boolean Satisfiability (SAT)~\cite{25} or Satisfiability Modulo Theories (SMT) solvers~\cite{69}. Besides, the nature of cryptographic algorithms involves math operations on a vast state space that makes it non-trivial for SMT solvers to cover~\cite{25}. Also, some cryptographic libraries rely on UNIX sockets and file operations to encode and decode text; BMC approaches require models for the environment. Besides, fuzzing often suffers from a low code coverage dilemma~\cite{51}. Also, fuzzing techniques face challenges to detect vulnerabilities in multi-threaded programs~\cite{87} since existing fuzzing techniques do not stress thread interleavings that affect execution states.

 We combine BMC with the fuzzing technique to reach parts of the code that BMC may fail to reach. The combination will also help the fuzzer reach deep parts of the code due to the initial seed generated from BMC. Also, We implement our custom LLVM pass to force the operating system to change the threads by injecting delay functions, thus stimulating the thread context switch. 
Here we develop a novel verification method named Encryption-BMC and Fuzzing (\tool), which exploits BMC and fuzzing to detect security vulnerabilities: memory corruption and concurrency vulnerabilities in IoT protocols' concurrent implementations. Note that client and server in communication protocols behave like different threads, and some encryption libraries are using POSIX threads in their implementation~\cite{94}. Although we focus on verifying cryptographic protocols, which represent a fundamental challenge due to the complex interleavings of client and server operations, our approach applies broadly to detecting vulnerabilities in concurrent implementations. In particular, we utilize BMC techniques to provide valuable seeds to our fuzzing approach to discover different thread interleavings, which make fuzzing detect vulnerabilities more efficiently~\cite{you2019slf}.

In contrast to Ognawala et al.~\cite{28}, who combines symbolic execution and fuzzing and apply it to general-purpose software, \tool starts with BMC. It then uses fuzzing by considering intricate security properties in IoT protocols since a server can produce multiple threads to accept requests from multiple clients simultaneously~\cite{yang2005performance}. We model (simulate) the client and server communication in IoT protocols as two threads and instrument the program to exploit different thread interleavings. \tool employs techniques, such as constant folding, bound $k$, and induction techniques, to reduce the number of states needed to be verified. Also, \tool relies on dynamic techniques such as fuzzing to explore the paths in the deployed environment that BMC requires models for.

%\subsection{Evaluation}
We evaluate \tool on five directories of the SV-COMP concurrency benchmarks.\footnote{\url{https://github.com/sosy-lab/sv-benchmarks}} We compare the results with \textit{ESBMC}~\cite{78}, \textit{AFL}~\cite{79}, Lazy Sequentialization (\textit{LazyCSeq})~\cite{95}, and Thread Sanitizer (\textit{TSAN})~\cite{91} on the same benchmarks. We also examine the \tool tool by verifying the WolfMQTT\cite{wolfmqtt} implementation. Moreover, we replicate known vulnerabilities in OpenSSL~\cite{openssl} and CyaSSL~\cite{cyassl}. The experimental results show that \tool outperforms the state-of-the-art verification tools as it detects vulnerabilities in more programs than the other tools.
In summary, we make the following two significant contributions towards the verification of cryptographic protocols targeted for IoT devices and concurrent implementations:
\begin{itemize}
  \item We propose a new hybrid verification method, named \tool, that combines BMC and fuzzing to increase code coverage and detect both memory corruption and concurrency vulnerabilities of IoT cryptographic protocols and concurrent implementations.
  \item We implement \tool to verify cryptographic IoT protocols and concurrent implementations. We show that \tool can find vulnerabilities which other existing tools, such as \textit{ESBMC}~\cite{78}, \textit{AFL}~\cite{79}, \textit{Lazy-CSeq}~\cite{95}, and \textit{TSAN}~\cite{91}, are unable to detect. \tool can also detect a data race bug in the open-source library WolfMQTT~\cite{wolfmqtt} and a thread leak in one example of WolfSSL implementation~\cite{94}. Also, \tool detects well known vulnerabilities in OpenSSL~\cite{openssl} and CyaSSL~\cite{cyassl}.
\end{itemize}
 
%=-=-=-=-=-=-=-=-=-=-=-=-=-=-=-=-=-=-=-=
\section{Verification Methods for Cryptographic Protocols and Concurrent Programs}
%=-=-=-=-=-=-=-=-=-=-=-=-=-=-=-=-=-=-=-=
Although the obligation to verify cryptographic protocols and concurrent programs is now well identified, only a few recent studies suggest solutions.

One of the attempts to verify cryptographic primitives using symbolic execution is suggested by Vanhoef and Piessens~\cite{25}. They modified the \textit{KLEE} tool~\cite{cadar2008klee} to efficiently handle cryptographic protocol by simulating their behavior under the Dolev-Yao model. Similarly, Given-Wilson et al.~\cite{98} proposed a process using model checking to detect fault injection vulnerabilities in the PRESENT cipher binary. The authors' framework used MC-Sema, which supports only some of the X86 architecture; they combined LLBMC with MC-Sema, which does not generate a trace to understand vulnerabilities and analyze results.

Another tool used for cryptographic primitives verification, based on fuzzing, is \textit{CDF}~\cite{56}. It is used to achieve security verification, and in particular, to find logic bugs with standard specifications. It uses differential fuzzing techniques to find inconsistencies between two implementations of the same primitive, e.g., of the RSA cipher~\cite{56}. \textit{CDF} is only effective when different implementations of the same algorithm are available, and these implementations do not contain the same bug. \textit{ESPIKE}~\cite{52} is another fuzzing tool, an extension of SPIKE~\cite{aitel2002introduction}, designed to handle secure protocols by sending all the \textit{SPIKE} data through the SSL layer~\cite{52}. Its limitation is that it is only valid for the already compatible protocols with \textit{SPIKE}.

Concerning concurrent programs, a few attempts have been proposed to detect security vulnerabilities. The challenge with a multi-threaded program is that it contains different thread interleavings, which may introduce bugs (e.g., data race) that are difficult to detect. \textit{MUZZ}~\cite{87} is a recent tool suggested for fuzzing concurrent programs. It is a grey box fuzzing tool that detects bugs in a multi-threaded program using thread-aware instrumentation. \textit{MUZZ}, similarly to \tool, instruments the code using LLVM pass to detect concurrency bugs. However, \tool uses a BMC technique to analyze the code and generate inputs, which can help the fuzzer trigger intricate execution paths as we show in \nameref{results} subsection. \textit{ConAFL}~\cite{92} is another thread-aware grey box fuzzer that focuses on user-space multi-threaded programs. It also uses heavy thread-aware static and dynamic analysis, which causes scalability issues. \textit{ConAFL} employs static analysis to locate sensitive concurrent operations to determine the execution order, focusing on three types of vulnerabilities: \textit{buffer-overflow}, \textit{double-free}, or \textit{use-after-free}. In contrast to \textit{ConAFL}, \tool is also able to detect more memory corruption bugs and concurrency bugs for maximum coverage.

%=-=-=-=-=-=-=-=-=-=-=-=-=-=-=-=-=-=-=-=
\section{\tool Design and Implementation}
%=-=-=-=-=-=-=-=-=-=-=-=-=-=-=-=-=-=-=-=

\begin{figure}[t!]
\centering
\includegraphics[width=\textwidth]{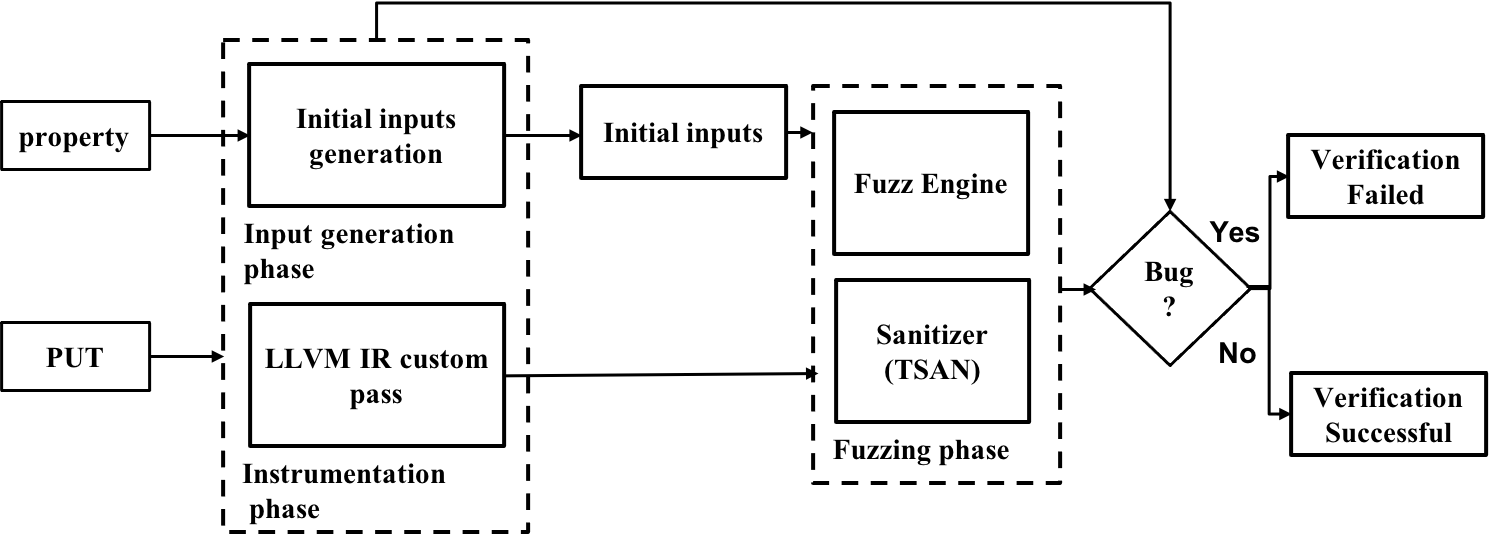}
\caption{Encryption-BMC and Fuzzing (\tool). }
\label{fig1}
\end{figure}

We develop a novel verification method to detect memory corruption vulnerabilities, such as buffer overflow and memory leak, and concurrency vulnerabilities, such as data races and thread leaks, using BMC and fuzzing techniques. We build the \tool verification method on top of two tools,~\textit{ESBMC} and \textit{AFL}, respectively; these tools have been chosen based on the comparison by Beyer et al.~\cite{73} between software verification and testing. 
Figure~\ref{fig1} illustrates the \tool verification method, which consists of three phases: \textit{input generation}, \textit{instrumentation}, and \textit{fuzzing}. Before describing these three phases, we first provide an overview of \tool.

%=====================================
\subsubsection{Overview.}
%=====================================

\tool can detect vulnerabilities in single C files. It uses \textit{ESBMC} for initial state exploration to search for different properties such as memory leaks and buffer overflows. We provide \textit{ESBMC} with the Program Under Test (PUT) and specific properties (P). If~\textit{ESBMC} detects a property violation and generates a counterexample, \tool extracts the assumption values and feeds them to the fuzzer as inputs to find unexpected paths that may expose a vulnerability. \textit{AFL} needs an initial seed to start fuzzing, and since \textit{AFL} is a smart fuzzer, it needs to understand the input type and mutate them to generate more seed inputs, which might trigger different paths. It is not implied that these inputs will generate the same violations as \textit{ESBMC}. It will only help the fuzzer observe different paths~\cite{chowdhury2019verifuzz}. In case~\textit{ESBMC} fails to detect a violation, then \tool generates random numbers to feed it to the fuzzer. 
As a second phase, we instrument the PUT using a custom LLVM pass to track the active threads and inject a delay function after each instruction at runtime. \tool feeds the inputs, and the LLVM pass to \textit{AFL} with \textit{TSAN's} help to search for additional bugs, especially concurrency bugs. Then, \tool analyzes the results; if there is a bug, it states~\textit{"Verification Failed"}; otherwise, it states~\textit{"Verification Successful"}. 

%======================================================  
\subsubsection{Input generation phase.}
\label{phase1}
%======================================================

This phase builds on top of \textit{ESBMC}. The user feeds \tool with the source code that needs to be tested with the specified proprieties, i.e., \texttt{unreach-call}, \texttt{valid-memsafety} and \texttt{no-overflow}, which is similar to SV-COMP~\cite{89}. When the user sets \texttt{unreach-call} property, it means there is a particular function call in the code that must be unreachable. When \texttt{valid-memsafety} is chosen, a specific memory safety property must hold in the code. Also, \texttt{no-overflow} propriety means there is a certain kind of undefined behaviour (i.e., overflows of signed integer) that must not exist in the code. Also, different properties will be checked in this phase for cryptographic protocol implementations. In particular, we check properties such as data races (\textit{--data-races-check}) and arithmetic overflow (\textit{--overflow-check}), which could expose vulnerabilities due to the communication between two entities. We also specify a bound $k$, using (\textit{--unwind k}) that will limit the visited areas of data structures such as arrays or the number of loop iterations.
\par Algorithm~\ref{alg:ESBMC}~\cite{CordeiroFM12} describes the BMC workflow inside the \tool verification method. Note that BMC take two parameters, the PUT and P, which need to be checked. From lines $1$ to $4$, BMC simplifies the PUT to a control flow (GOTO program). Then, it converts the GOTO program to a single static assignment (SSA) form. In line $5$, it converts SSA into quantifier-free formula $(C \wedge \neg P)$, where $C$ denotes constraints and $P$ denotes properties. Then, the SMT solver checks the formula satisfiability. From lines $7$ to $10$, BMC checks if $T$ is satisfiable; if so, it converts the result into a counterexample. If the result is unsatisfiable, then it returns verification successfully.
However, loop unwinding is essential for the BMC technique, as it is responsible for constraining the state space that the algorithm explores~\cite{gadelha2020esbmc}. When using BMC, we specify a bound $k$, which will limit the visited
areas of loop iterations. This limits the state space to be explored during verification, leaving enough time to find real errors in programs. \par On the one hand, if \textit{ESBMC} finds a violation of the properties and generates a counterexample. \tool extracts the assumptions values from the counterexamples and saves them for the fuzzer. On the other hand, when \textit{ESBMC} fails to detect a violation, \tool then generates random numbers ranging between $0$ to $300$ as an input to the fuzzer. We set the range to small numbers as we observed that large numbers, for example, more than $500$ will make the compiled program slow or hang for fuzzing. We also observed that these numbers provide a good trade-off between functionality and efficiency -- \tool detects more bugs in less time.

  \begin{algorithm}
\caption{BMC}\label{alg:ESBMC}
\textbf{Input:} Program Under Test (PUT) and proprieties (P) to be checked.\\
\textbf{Output:} FAIL (SMT) with counterexample (propriety violation detected by ESBMC).
\begin{algorithmic}[1]
\State $PUT \gets$ Specify source code
\State $P \gets$ Specify proprieties 
\State $C \gets$ Convert $(PUT)$ to control flow
\State $S \gets$ Convert(C) into (SSA) form 
\State $E \gets $ Convert (S) into quantifier free formula $(C \wedge \neg P)$
\State $T \gets$ Solve (E) with SMT solver
\If {$T$ satisfiable}
\State Convert(T) into counterexample
\Else
\State  \textbf{return} Verification Successful;
\EndIf
\end{algorithmic}
\end{algorithm}

%--------------------------------------------------------------
\subsubsection{Instrumentation phase.}
\label{phase2}
%--------------------------------------------------------------

In this phase, we instrument the PUT to compile it with the techniques explained in the~\nameref{phase3}. To achieve this, we developed a custom LLVM Pass~\cite{97} using LLVM version 10, which is used to perform the transformations and optimizations in the program. Specifically, \tool instruments the PUT by injecting a random delay function after each instruction at the LLVM intermediate representation level (LLVM-IR) by changing the IR itself and not the original source code. We implement the functions (e.g., \texttt{delay}, \texttt{Pthread\_add}, and \texttt{Pthread\_release}) as a run-time library in C and link it with the LLVM pass at compile time. The delay function is a nanosecond sleeping function, ranging from $1$ to $100,000$ nanoseconds. This specific time does not add an overhead to the binary. As we observed, increasing the time of the delay will make the fuzzer abort. However, the LLVM pass keeps tracking the active threads by countering~\textit{Pthread\_create} and switches to the delay function to be executed. Note that if it encounters~\textit{Pthread\_join}, no active threads are running, then it switches to no delay. This approach keeps the instrumentation lightweight and helps the fuzzer detect any unsynchronization between threads. Therefore, with the appropriate inputs, our fuzzer can detect vulnerabilities in concurrent programs. 
Listing~\ref{lst:1} illustrates an overview of how our LLVM pass works. In lines $4, 7$ and $10$, we insert the delay function after each LLVM instruction. Then, in line $4$, we encounter the~\textit{Pthread\_create} function, which means there is a live thread; the LLVM pass will insert \textit{pthread\_add} function from the run-time library. After that, in line $7$, we run the delay function. Then, in line $8$, we encounter~\textit{Pthread\_join}, which will mark the thread as not live by inserting the function \textit{Pthread\_release} from the runtime library. After that, in line $10$, the delay function will not run the delay.

 \begin{lstlisting}[language=python, caption=Overview of how the LLVM pass works.,label={lst:1}]
void main()
{
Some statements,
_delay_function() #inserted but do not delays
pthread_create(&t1, 0, thread1, 0); #one thread is created, pthread_add is called 
Some statements,
_delay_function()  #inserted and do real delays 
pthread_join(t1, 0); #the thread is not live, pthread_release is called
Some statements,
_delay_function()  #inserted but do not delay again
}\end{lstlisting}
 
Algorithm \ref{alg:pass} illustrates the steps behind the LLVM pass. From lines $1$ to $3$, it iterates over each Function's Basic Blocks $BB$, then after each instruction $I$ inside $BB$, a call to the delay function $D_f$ is inserted. From lines $5$ to $9$, the pass iterates over each Function's Basic Blocks; then, if it encounters a call to~\textit{Pthread\_create}, it inserts a call to a function \textit{Pthread\_add}, whose purpose is to count the active threads and then switch delay function to continue to do the delay. In lines $10$ to $12$, if it encounters a call to~\textit{Pthread\_join}, it inserts a call to a function \textit{Pthread\_release}, which reduces the number of active threads by one and switches the delay function to return without delay.

\begin{algorithm}
\caption{LLVM Custom Pass}\label{alg:pass}
\textbf{Input:} Program Under Test (PUT).\\
\textbf{Output:} Instrumented program.\\
\textbf{INITIALIZE FUNCTIONS:} (Delay function $(D_f)$, pthread\_add function $(A_f)$ and pthread\_release function$(R_f)$).\\
\textbf{INITIALIZE TARGET:}(pthread\_create  and  pthread\_join).
\begin{algorithmic}[1]
\ForAll{Function $F \in PUT$}
\For{Basic blocks $BB$ in $F$}
 \If{$I$ is a branch Instruction }
 \State $M \gets$ insert $D_f$ \Comment{insert a call to delay function after each instruction,}
 \EndIf
 \EndFor
    \EndFor
\ForAll{Function $F \in PUT$} 
\For{Basic blocks $BB$ in $F$}
\If{$I$ == pthread$\_$create} \Comment{Track each pthread$\_$create}
\State $C \gets$ insert $A_f$ \Comment{Count as an active thread}
\State switch $D_f$ to delay \Comment{ Continue to run delay}
\ElsIf{$I$ == pthread$\_$join} \Comment{Track each pthread$\_$join}
\State $J \gets R_f$ \Comment{Count as NOT an active thread}
\State switch $D_f$ return \Comment{No delay's running }
\EndIf
 \EndFor
    \EndFor
\end{algorithmic}
\end{algorithm}

%----------------------------------------------
\subsubsection{Fuzzing phase.}

\label{phase3}
%----------------------------------------------
In this phase, we build our fuzzing engine on top of \textit{AFL} and \textit{TSAN}. \tool feeds \textit{AFL} with the inputs generated from~\nameref{phase1} and the LLVM pass generated from~\nameref{phase2} with the help of \textit{TSAN}. Specifically, we compile the source code itself (and not the binary) with the LLVM pass using the \textit{AFL} clang wrapper~\cite{79} and utilize the Thread Sanitizer (TSAN) flag.
\textit{AFL} then feeds the code with mutation inputs to execute different paths.
Algorithm~\ref{alg:AFL}~\cite{lemieux2018fairfuzz,87} shows the standard workflow of a grey-box fuzzer (GBF) such as \textit{AFL}~\cite{79}. 
A GBF takes a target program PUT and initial seeds $M$ as inputs. It uses its instrumentation to track code coverage $P_f$ and then starts the loop in line $1$; from lines $3$ to $5$, it selects the seeds and schedules them by applying the same number of mutations $N$ that are applied to $t$ to generate the mutated seed $t'$. GBF mutates inputs in two main stages: the \textit{Deterministic} and the \textit{Havoc}~\cite{lemieux2018fairfuzz}. On the one hand, all the deterministic mutation stages work by traversing the input under mutation and applying a mutation at each input position. These mutations include bit flipping, arithmetic increment and decrement of integer values, etc.
On the other hand, the havoc stage operates by applying a sequence of random mutation, setting random bytes to random values and deleting subsequences of the input to the input being mutated to create a new input. These mutation strategies assume that the input to the PUT is a sequence of bytes. The mutated inputs produced on each of these stages are governed by the length of the input being mutated.
From line $6$, the fuzzer repeatedly executes $N$ times, for each new seed $t'$, to get the execution statistics. In line $9$, the inputs from $t'$ are evaluated based on the statistics and coverage feedback from the instrumentation $p_f$. If the input triggers a crash, it saves it in the crash directory and marks it as a unique crash or if it covers a new branch, it saves it in the seed queue.

\begin{algorithm}
\caption{Grey-box Fuzzing}\label{alg:AFL}
\textbf{Input:} Program Under Test (PUT) and CORPUS directory that contains the initial seeds (M).\\
\textbf{Output:} final seed queue $(Q_S)$, vulnerable inputs file $(S_I)$.
\begin{algorithmic}[1]
\State $P_f\gets$ instrument $(PUT)$  \Comment{Fuzzer instrument the source code}
\State $S_I \gets$ $\phi $
\While {true}
\State $t \gets$ select next seed $(Q_S)$ \Comment{Seed selection}
\State $N \gets$ get mutation chance $(P_f, t)$\Comment{Seed scheduling}
\ForAll{$i \in 1 \cdots N$}
    \State $t' \gets$ mutated input $(t)$ \Comment{Seed mutation}
    \State $rep \gets$ Run $(P_f, t’, M_c)$ \Comment{Repeated execution}
    \If{is crash$(rep)$} 
      \State $S_I \gets$ $S_I \cup {t'}$\Comment{Vulnerable seed} 
    \ElsIf {cover new trace $(t’, rep)$}\Comment{Maintain “effective” seeds}
      \State {$Q_i \gets Q_i \oplus t'$}
    \EndIf
\EndFor

\EndWhile
\end{algorithmic}
\end{algorithm}

In \tool, the mutation and next seed scheduling can generally affect the results because the mutated seed inputs can achieve new coverage. It can detect vulnerabilities in the new execution path, which \tool will report.
In this phase, we aim to detect and report the memory corruption errors in concurrent programs such as buffer overflow and memory leak using \textit{AFL} and detecting concurrency bugs such as data race and thread leak using \textit{TSAN}. However, \textit{AFL} performance suffers because it does not track all possible schedule interleavings~\cite{87}. To overcome this limitation, we inject random delays in~\nameref{phase2} to help \textit{AFL} detect different thread interleavings. This way, it has more possibilities to detect concurrency bugs, as demonstrated in our experimental evaluation.
In specific, since \textit{AFL} is not aware of threads interleaving~\cite{87}, we instrument the source code to force the operating system to change the thread execution by injecting delay functions, thus stimulating the thread context switch. This increases the chances of detecting bugs in concurrent implementations.
In the end, \tool analyzes the results and generates the bug report with all the bugs detected, which are either memory corruption bugs or concurrency bugs.

%=-=-=-=-=-=-=-=-=-=-=-=-=-=-=-=-=-=-=-=
\section{Experimental Evaluation}
%=-=-=-=-=-=-=-=-=-=-=-=-=-=-=-=-=-=-=-=

%=-=-=-=-=-=-=-=-=-=-=-=-=-=-=-=-=-=-=-=
\subsection{Description of the benchmarks and setup}
%=-=-=-=-=-=-=-=-=-=-=-=-=-=-=-=-=-=-=-=

We build the \tool tool using python and C++ programming languages. We evaluate it over various benchmarks, specifically the open-source implementation WolfMQTT~\cite{wolfmqtt} and SV-COMP directories: \texttt{Pthread}, \texttt{Pthread-atomic}, \texttt{Pthread-divine},  \texttt{Pthread-complex}, and  \texttt{Pthread-lit} from the concurrency safety category~\cite{96}, which includes $81$ verification tasks. We also replicate some known vulnerabilities in OpenSSL and CyaSSL. %\textcolor{red}{I already added them in the introduction, do you want me to remove the citation from there and put it her? }
Here we give a brief overview of the tools we used in our comparison. \textit{AFL} is a GBF that uses evolutionary genetic algorithms and runtime instrumentation to discover new interesting inputs that trigger new internal states in the targeted binary~\cite{you2019slf}. \textit{ESBMC} is a context-bounded model checker based on SMT solvers to verify single and multi-threaded C/C++ programs~\cite{gadelha2020esbmc}. \textit{TSAN} is a data race detector for C/C++ programs~\cite{91}. It employs compile-time instrumentation to examine all non-race-free memory access at runtime. \textit{Lazy-CSeq} is a context-bounded verification tool that translates a multi-threaded C program into a sequential consistent C program~\cite{95}. We chose these tools for evaluation even when most of them are actively used in \tool to prove that \tool is outperforming the other by taking each approach's advantages, notably, because of our evolutionary custom LLVM pass.

We run \tool on each of these $81$ verification tasks and compare its results to \textit{ESBMC} v6.4.0, \textit{Lazy-CSeq} v2.1, \textit{AFL} v2.5b, and \textit{TSAN} clang version 10.0.0. We set \texttt{unreach\_call} property for all the verification tasks since the concurrency safety category required this specification. However, we set some different properties for cryptographic protocol implementations that used in ESBMC such as \texttt{--data-races-check} and \texttt{--overflow-check}. This helps ESBMC to detect violations of concurrency and security. Regarding \textit{AFL}, we set the \textit{TSAN} flag to detect concurrency bugs. 
For evaluation, we used three metrics: the number of bugs detected, the time consumption and the memory used.
All experiments were conducted on an idle Intel Core i7 2.7Ghz processor, with 8 GB of RAM and running Ubuntu 18.04.5 LTS. All
tools, benchmarks and results of our evaluation are available on GitHub.\footnote{\url{https://github.com/fatimahkj/EBF}}

%=-=-=-=-=-=-=-=-=-=-=-=-=-=-=-=-=-=-=-=
\subsection{Goals}
%=-=-=-=-=-=-=-=-=-=-=-=-=-=-=-=-=-=-=-=

Our main experimental goal is to check the \tool method's effectiveness and performance to verify cryptographic protocols, mostly concurrent implementations of such protocols. Our experimental evaluation has the following two goals:
\begin{tcolorbox}[width=\textwidth]
\begin{enumerate}
\item[EG1] \textbf{Concurrent bug detection:} To demonstrate that \tool can detect more bugs in multi-threaded programs than other state-of-the-art verifiers.
\item[EG2] \textbf{Cryptographic protocol bug detection:} To demonstrate that \tool can be employed to detect bugs in real-world cryptographic protocols.
\end{enumerate}
\end{tcolorbox}

%=-=-=-=-=-=-=-=-=-=-=-=-=-=-=-=-=-=-=-=
\subsection{Results}
\label{results}
%=-=-=-=-=-=-=-=-=-=-=-=-=-=-=-=-=-=-=-=

%-------------------------------------------
\subsubsection{SV-COMP benchmarks.}
%-------------------------------------------
The SV-COMP benchmarks provide a wide range of verification tasks that evaluate the verifiers' different strengths, including some TLS examples. 
We run \textit{ESBMC}, \textit{AFL}, \textit{TSAN} and \textit{Lazy-CSeq} in isolation with the same benchmarks using clang v10 in our local machine except for \textit{Lazy-CSeq}, where we use the results reported in SV-COMP 2021. Then, we compare the results with \tool.
\begin{table}[t!]
\centering
\caption{Experimental results of the comparison between \tool and \textit{ESBMC},~\textit{AFL}, \textit{TSAN} and \textit{Lazy-CSeq}. It consists of the main category denoted by ``Concurrency Safety'' with five different directories (\textit{Pthread}, \textit{Pthread-atomic}, \textit{Pthread-divine}, \textit{Pthread-complex} and \textit{Pthread-lit}); each directory consists of different number of tasks with total line of code (LOC). The number of tasks each tool found a bug is represented for each directory (e.g., $19$ tasks out of $38$ \textit{\tool} found a bug in them)}\label{tab1}
\begin{tabular}{|l|c|c|c|c|c|c|c|}
\hline
\textbf{Concurrency Safety} &  \textbf{No. of asks} & \textbf{LOC} & \textbf{ESBMC} & \textbf{AFL} & \textbf{TSAN} & \textbf{Lazy-CSeq} & \tool\\
\hline \hline
Pthread  &  38 & 3674 &8 & 4 & 17 &15 & \textbf{19}\\
\hline
Pthread-atomic &  11 & 1669 &0 & 2 & \textbf{6} & 2 & \textbf{6}\\
\hline
Pthread-divine &  16 &333 &2 & 0 & 13 & 7 & \textbf{14}\\
\hline
Pthread-complex & 5 & 1633 & 2 & 0 & \textbf{3} & 2 & \textbf{3}\\
\hline
Pthread-lit & 11 & 595 & 1 & 1 & 8 & 3 & \textbf{9}\\
\hline
Total &  81 & 7904 & 11 &7 & 47 & 29& \textbf{51}\\
\hline
\end{tabular}
\newline
\end{table}
In this evaluation, we check how many bugs were detected with \tool compared with the other tools. 

Table~\ref{tab1} illustrates the five directories (e.g., \texttt{Pthread}, \texttt{Pthread-atomics}, \texttt{Pthread-divine}, \texttt{Pthread-complex}, and \texttt{Pthread-lit}) with the number of verification tasks (benchmarks) and lines of code (LOC). Bold numbers indicate the best results for each directory (number of benchmarks in which we discovered vulnerabilities). It also shows the tools evaluated and compared with \tool; note that~\textit{Lazy-CSeq} is the SV-COMP 2021 winner. For completeness, in the~\nameref{phase1}, after several experiments on \tool, we noticed that setting the unwinding bound limits to $k=20$ is the best bound for evaluating concurrency safety benchmarks, which mitigates time and memory exhaustion with a suitable value to expose a vulnerability. Therefore, in \tool, we increment the loop unwinding until it reaches $k=20$, to mitigate exhaustion of time or memory limits for programs with loops whose bounds are too large. 

Overall, \tool outperforms all evaluated tools. For the \texttt{Pthread} directory that consists of $38$ tasks, \tool significantly outperform \textit{ESBMC}, \textit{AFL}, \textit{Lazy-CSeq}, and \textit{TSAN} by two verification tasks. \tool also outperforms all evaluated tools in the \texttt{Pthread-divine} and \texttt{Pthread-lit} directories, and \textit{TSAN} by one verification task in each directory. However, in the \texttt{Pthread-atomic} and \texttt{Pthread-complex} directories, \tool detected a bug in the same tasks as \textit{TSAN}, but it outperforms all other evaluated tools.

\begin{figure}[t!]
\centering
\includegraphics[width=\textwidth]{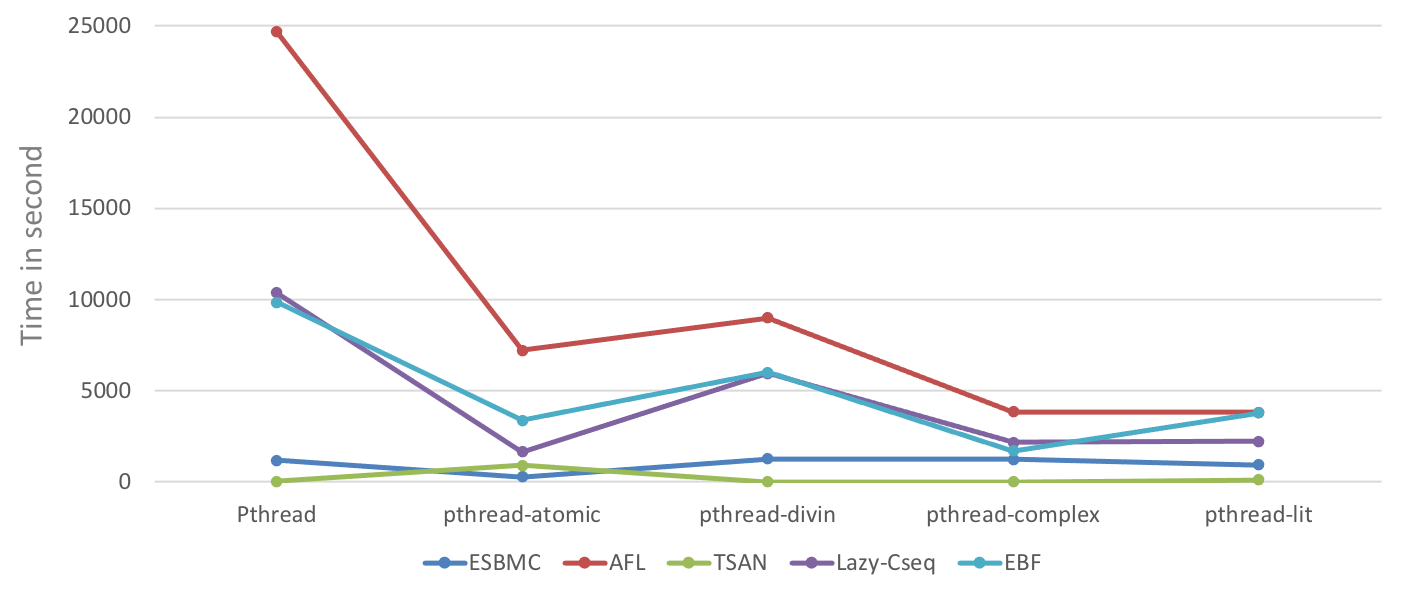}
\caption{Total time consumption in second for each directory}
\label{fig2}
\end{figure}

In \tool, we focused on reporting the number of benchmarks in which we detected a bug compromising the time and memory it takes. We set the time limit for each tool to $15$ minutes. To achieve equality with \textit{Lazy-CSeq} that sets the time in SV-COMP results to 15 minutes. 
However, Table~\ref{tab2} shows the total time it took for the tools to detect vulnerabilities in all five directories. \tool was not the best tool regarding the time. It takes a total time of around $24657$ seconds, which consumes more time than \textit{ESBMS, TSAN} and \textit{Lazy-CSeq}. Figure~\ref{fig2} shows the total time for each tool for each directory. \textit{AFL} takes more time than the other tools in all directories. \tool and \textit{Lazy-CSeq} compete in second and third place in terms of efficiency. \tool takes less time than \textit{Lazy-CSeq} in \texttt{Pthread} and~\texttt{Pthread-complex}, and almost the same time in \texttt{Pthread-divine}. %\textcolor{red}{MM: Bring Fig 2 and Table 2 closer to this paragraph.}
\begin{table}[t!]
\centering
\caption{Total time consumption in second.} %\textcolor{red}{MM: Why not have again the text in bold for the best tool - the one that takes the minimum time?} }
\label{tab2}
\begin{tabular}{|l|c|c|c|c|c|c|}
\hline
\textbf{Tools} & \textbf{ESBMC} & \textbf{AFL} & \textbf{TSAN} & \textbf{Lazy-CSeq} & \tool\\
\hline \hline
Total Time (in sec) &  4838 & 48518 & \textbf{991} &  22314 & 24657 \\
\hline
\end{tabular}
\newline
\end{table}

\begin{figure}[t!]
\centering
\includegraphics[width=\textwidth]{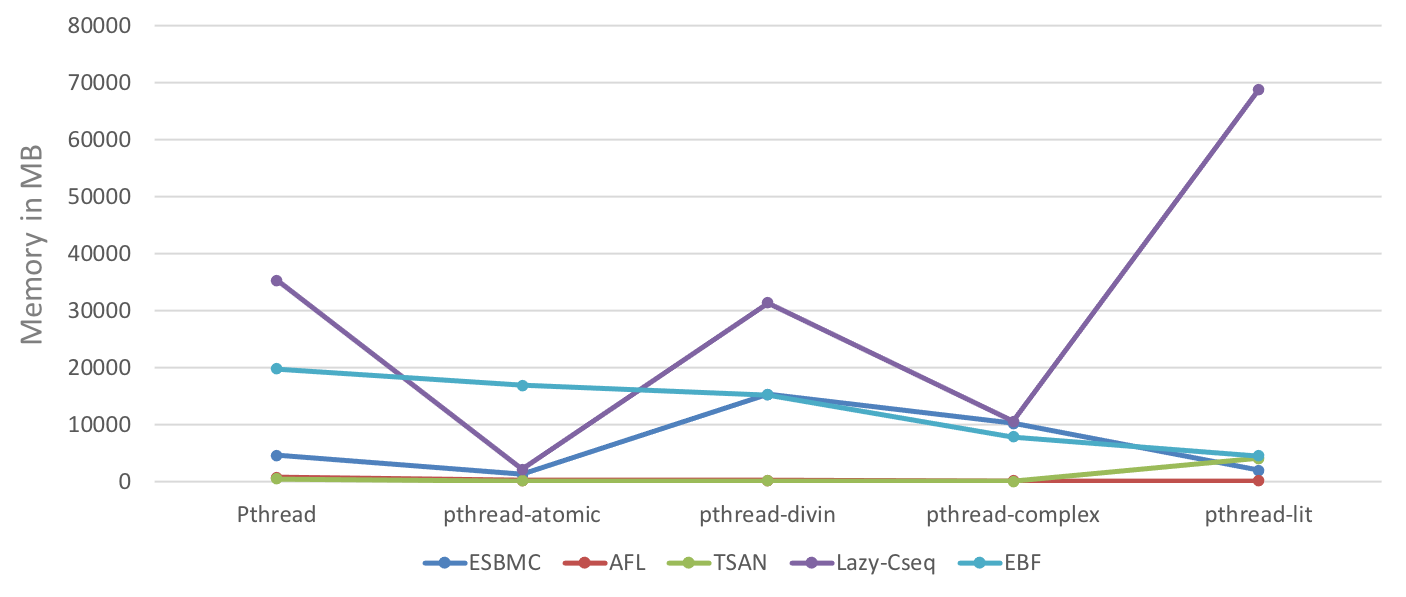}
st    \caption{Memory consumption for each directories in MB}
\label{memoryeach}
\end{figure}

Regarding memory consumption, Figure~\ref{memoryeach} shows the memory consumption for each directory expressed in MB; for all the directories, \textit{AFL} consumes the least memory. \textit{TSAN} consume less memory steadily in all the directories except in \texttt{Pthread-lit}. For \textit{Lazy-CSeq}, it fluctuates between the directories, while \tool consumes higher memory in the \texttt{Pthread} directory and descends gradually. Lastly, \textit{ESBMC}, consume less memory than \tool except in \texttt{Pthread-complex}.

Note that none of the evaluated tools is consistently better than the other. Each verifier has advantages and disadvantages. The effectiveness 
of \tool depends mainly on its capability of detecting both memory corruption and concurrency bugs, which maximize the chance of finding more bugs. Therefore, we compromise the time and memory limits for the capability of detecting more vulnerabilities. \textit{ESBMC} can find both types of vulnerabilities except that it does not perform well in concurrent implementations, as shown in Table~\ref{tab1}. \textit{ESBMC} either detects a violation or exhausts time or memory limits, while \textit{Lazy-CSeq} performed well except in \texttt{Pthread-lit} directory. Note that \textit{Lazy-CSeq} v2.1 does not check for data races, which makes its performance worse compared to \tool and \textit{TSAN}. Also, since \textit{AFL} is not effectively detecting concurrency vulnerabilities, it performed worse than all other tools in most directories. Lastly, \textit{TSAN} performs relatively well in most directories compared to the other tools.

\begin{tcolorbox}[width=\textwidth]
Overall, \tool detected bugs in 51 out of 81, with four verification tasks more in total than \textit{TSAN} (the best competing tool). \tool shows encouraging results compared with the state-of-the-art software testing and verification tools, which answers~\textbf{EG1}. 
\end{tcolorbox}
\subsubsection{WolfMQTT verification.}
To verify the WolfMQTT implementation, we used \tool over its API functions specifically, ``MqttSocket\_Write'' and ``MqttSocket\_Read'' by verifying different files as~\texttt{multi\-thread.c}, which contains a set of tests for WolfMqtt in multi-threads contexts. \tool could detect a data race in the ``MqttClient\_WaitType'' function. To compile the source code, we execute this experiment with a Mosquito server~\cite{mosquitto} running on the same machine as the client. As part of our research effort, we raised a GitHub issue to warn the developers of this potential bug. The time it takes for \tool to detect the data race was $15$ min with 23.228 MB memory consumption. We run \textit{ESBMC} and \textit{AFL} on WolfMQTT, but they did not detect the bug. \tool detected the bug with our evolutionary instrumentation.\footnote{\url{https://github.com/wolfSSL/wolfMQTT/issues/198}} We also run \tool over an example of WolfSSL. \tool detected a thread leak in \texttt{memory-tls.c}, which was reported in GitHub as an issue to the WolfSSL developers.\footnote{\url{https://github.com/wolfSSL/wolfssl-examples/issues/242}} A thread leak happens when the created thread is not released or joined, potentially causing the application to be slow or hang. However, even if the WolfSSL is a simple demonstration, not an actual application, it should be secured.
\subsubsection{Known vulnerabilities.}
Furthermore, to show \tool effectiveness, we replicate some known vulnerabilities with assigned CVE number and verify them with \tool; the first one is OpenSSL v 1.0.1, with $CVE-2014-3512$, this vulnerability occurs in ``srp\_lib.c``, which has buffer overflow as described in the CVE website~\cite{2014-3512}. We replicated the pointer dereference issue that is happening when calling the methods without properly initialized pointers. The pointer dereference occurs at the function``SRP\_calc\_u`` in ``srp$/$srp\_lib.c``. \tool could detect this vulnerability in $302$ seconds and used $121$ MB of memory. The second known vulnerability is also in OpenSSL v 1.0.1; the assigned CVE number is $CVE-2016-2106$~\cite{2016-2106}. An integer overflow occurs in the ``EVP\_EncryptUpdate`` function in ``crypto$/$evp$/$evp\_enc.c". \tool was able to replicate it directly by creating a test file that calls the ``EVP\_EncryptUpdate`` function with non-deterministic inputs and using \textit{--overflow-check} propriety inside \textit{ESBMC}. \tool detects the overflow in $301$ seconds and used $78$ MB of memory. The last vulnerability is out\-of\-bounds in the ``DoAlert`` function in CyaSSL v 2.9.0, which is later known as WolfSSL; the CVE number is $CVE-2014-2896$~\cite{2014-2896}. \tool verifies ``src/internal.c``, which contains ``DoAlert``. The ``DoAlert`` function is not checking the index limits when receiving a number pointer, causing the out-of-bound vulnerability. \tool detects the variability in $301$ seconds and used $529$ MB of memory.

%will we replicate the bugs we found bugs in ESBMC, which we report to the developer
%we compromise time and memory for a number of bugs detected
\begin{tcolorbox}[width=\textwidth]
\tool could detect a data race in one open-source implementation for cryptographic protocols, which thus answers \textbf{EG2}. 
\end{tcolorbox}

\begin{comment}

\begin{table}[!t]
\caption{Average time consumption in second for each directory. }
\centering
\label{Tab3}
\begin{tabular}{|l| c | c | c | c | c |} 
\hline
\multicolumn{1}{| c |}{\textbf{Concurrency Safety}} & \multicolumn{1}{| c |}{\textbf{Pt}}  & \multicolumn{1}{ c |}{\textbf{Pt-atomic}} & \multicolumn{1}{ c |}{\textbf{Pt-divine}}  & \multicolumn{1}{ c |}{\textbf{Pt-complex}}& \multicolumn{1}{ c |}{\textbf{Pt-lit}}\\ 
%\cline{2-11}
%\textbf{Tools} &\textbf{Total Time}  \\
\hline
\hline
 \textbf{ESBMC}   & 31  & 25 &   79 &  243 &  83     \\ \hline
 \textbf{AFL}   & 667 &  655 &  563 &  768 &  345     \\ \hline
 \textbf{TSAN}  &  $<$1 &  82 &  $<$1  & $<$1  &   08   \\ \hline
 \textbf{Lazy-CSeq}  & 273  & 149 &  371 &  432 &  202    \\ \hline
 \tool   & 259  & 307 & 375 & 337 &  342   \\ \hline
\end{tabular}
\end{table}
\end{comment}

%=-=-=-=-=-=-=-=-=-=-=-=-=-=-=-=-=-=-=-=
\subsection{Threats to Validity}
%=-=-=-=-=-=-=-=-=-=-=-=-=-=-=-=-=-=-=-=

We selected our SV-COMP benchmarks based on two factors. First, it contains a variety of vulnerable programs. Second, it employs POSIX threads. However, SV-COMP benchmarks contain specific functions, which are not following the C standard.~\footnote{\url{https://github.com/sosy-lab/sv-benchmarks/issues/1291}} For example, we excluded some directories such as ``Pthread-driver-races'' because some of the benchmarks contain compilation errors using clang 10 (e.g., unknown type name), and fixing these errors might introduce other bugs. As a result, we evaluated \tool on $81$ verification tasks from the concurrency safety category.
The main threat is that this paper's evaluation is subject to these benchmarks and may not be generalizable to other benchmarks. Another threat to our experiment's validity is that we evaluated our method on different examples of the WolfMQTT implementation. Still, those examples invoke several C functions and also one instance of the WolfSSL library. We also evaluated our method on several parts of OpenSSL v 1.0.1, not the entire library. In particular, \tool may work on a broader set of protocols. However, we did not evaluate \tool over such protocols as \tool was developed to work specifically on lightweight cryptographic protocol implementations. In general, \tool can be further developed to be effective on all the implementation differences of cryptographic protocols, but here we have evaluated it only on lightweight cryptographic protocol implementations.

%=-=-=-=-=-=-=-=-=-=-=-=-=-=-=-=-=-=-=-=
\section{Conclusions and Future Work}
%=-=-=-=-=-=-=-=-=-=-=-=-=-=-=-=-=-=-=-=

This paper presented \tool, a novel software verification tool that combines BMC and fuzzing techniques to detect memory corruption and concurrency vulnerabilities in IoT cryptographic protocols and concurrent implementations. \tool works by feeding \textit{AFL} with an initial seed generated from \textit{ESBMC} counterexamples along with our custom-developed LLVM pass. We run \tool over several SV-COMP benchmarks and an open-source application. We show that \tool outperforms other state-of-the-art software testing and verification tools such as \textit{ESBMC}, \textit{AFL}, \textit{Lazy\_CSeq} and \textit{TSAN} in detecting vulnerabilities in more tasks. We also show that \tool detected a data race in WolfMQTT and could detect known vulnerabilities in OpenSSL and CyaSSL. Thus, \tool contributes to the vision of fully verified trustworthy software systems. For future work, we plan to extend \tool to support more cryptographic protocol implementations and improve its performance by boosting the verification time.
%\textcolor{red}{MM: Done! Happy to have another read once you address these comments. }

% the environments 'definition', 'lemma', 'proposition', 'corollary',
% 'remark', and 'example' are defined in the LLNCS documentclass as well.
\bibliographystyle{splncs04}
\bibliography{mybibliography}
%

\begin{comment}

\end{comment}
\end{document}